\documentclass[letterpaper, 10 pt, conference]{ieeeconf} 

\IEEEoverridecommandlockouts               

\usepackage{graphics} 
\usepackage{epsfig} 
\usepackage{times} 
\usepackage{amsmath} 
\usepackage{amssymb} 
\usepackage{pbox}    

\usepackage{tikz}
\usepackage{pgfplots}
\pgfplotsset{compat=1.17}

\usepackage{subfig}
\usepackage{cite}
\usepackage{bm} 
\usepackage{xfrac}

\newtheorem{remark}{Remark}

\usepackage{algorithm}
\usepackage[noend]{algpseudocode} 

\usepackage{tikz}
\usepackage{pgfplots}

\definecolor{orange}{RGB}{217,83,25}
\definecolor{blue}{RGB}{0,114,189}
\definecolor{yellow}{RGB}{237,177,32}
\definecolor{purple}{RGB}{126,47,142}
\definecolor{green}{RGB}{119,172,48}

\newcommand{\Cline}[2]{\raisebox{2pt}{\tikz{\draw[#1,#2,line width=2pt](0,0) -- (5mm,0);}}}

\usepackage{setspace}
\makeatletter
\newcommand\fs@spaceruled{\def\@fs@cfont{\bfseries}\let\@fs@capt\floatc@ruled
	\def\@fs@pre{\vspace{0.6\baselineskip}\hrule height.8pt depth0pt \kern2pt}%
	\def\@fs@post{\kern2pt\hrule\relax}%
	\def\@fs@mid{\kern2pt\hrule\kern2pt}%
	\let\@fs@iftopcapt\iftrue}
\makeatother

\usepackage{textcomp}
\usepackage{hyperref}
\usepackage{lipsum}
\newcommand\copyrighttext{
	\footnotesize \textcopyright 2023 IEEE.  Personal use of this material is permitted.  Permission from IEEE must be obtained for all other uses, in any current or future media, including reprinting/republishing this material for advertising or promotional purposes, creating new collective works, for resale or redistribution to servers or lists, or reuse of any copyrighted component of this work in other works. DOI: \href{https://doi.org/10.1109/CDC49753.2023.10383769}{10.1109/CDC49753.2023.10383769}}
\newcommand\copyrightnotice{
	\begin{tikzpicture}[remember picture,overlay]
		\node[anchor=south,yshift=10pt] at (current page.south) {\fbox{\parbox{\dimexpr\textwidth-\fboxsep-\fboxrule\relax}{\copyrighttext}}};
	\end{tikzpicture}%
}

\title{\LARGE \bf Accelerating soft-constrained MPC for linear systems\\ through online constraint removal}

\author{S.A.N. Nouwens$^{1}$, M.M. Paulides$^{2,3}$, W.P.M.H. Heemels$^{1}$
\thanks{This research is supported by KWF Kankerbestrijding and NWO Domain AES, as part of their joint strategic research programme: Technology for Oncology II, Learn-2-Act (17918). The collaboration project is co-funded by the PPP Allowance made available by Health$\sim$Holland, Top Sector Life Sciences \& Health, to stimulate public-private partnerships. }
\thanks{$^{1}$Control Systems Technology section, Department of Mechanical Engineering, Eindhoven University of Technology, Eindhoven, The Netherlands
}%
\thanks{$^{2}$Electromagnetics for Care \& Cure, Department of Electrical Engineering, Eindhoven University of Technology, Eindhoven, The Netherlands}%
\thanks{$^{3}$Department of	Radiotherapy, Erasmus University Medical Center Cancer Institute, Rotterdam, The Netherlands}%
}

\begin{document}
\bstctlcite{IEEEexample:BSTcontrol} 

\maketitle
\thispagestyle{empty}
\pagestyle{empty}

\copyrightnotice
\begin{abstract}
	Optimization-based controllers, such as Model Predictive Control (MPC), have attracted significant research interest due to their intuitive concept, constraint handling capabilities, and natural application to multi-input multi-output systems. However, the computational complexity of solving a receding horizon problem at each time step remains a
	 challenge for the deployment of MPC. This is particularly the case for systems constrained by many inequalities. Recently, we introduced the concept of constraint-adaptive MPC (ca-MPC) to address this challenge for linear systems with hard constraints. In ca-MPC, at each time step, a subset of the constraints is removed from the optimization problem, thereby accelerating the optimization procedure, while resulting in identical closed-loop behavior. The present paper extends this framework to soft-constrained MPC by detecting and removing constraints based on sub-optimal predicted input sequences, which is rather easy for soft-constrained MPC due to the receding horizon principle and the inclusion of slack variables. We will translate these new ideas explicitly to an offset-free output tracking problem. The effectiveness of these ideas is demonstrated on a two-dimensional thermal transport model, showing a three order of magnitude improvement in online computational time of the MPC scheme.  	 
\end{abstract}

\vspace{-0.4em}
\section{Introduction}\label{sec:introduction}
Model predictive control (MPC) is a highly successful control technology in many industrial domains. Its strengths are, amongst others, the natural ability of constraint handling and straightforward application to multi-input multi-output systems. To implement this control technology, generally, MPC requires the solution of a finite horizon optimal control problem to be computed online. Despite MPC's success, for complex systems, the computational complexity of solving an optimization problem online can be a bottleneck for its real-time implementation. For this reason, there is a persistent pursuit to accelerate solving the online optimization problem. Approaches to accelerate MPC include, amongst others, advances in tailored solvers \cite{Frison2020Jan, Arnstrom2022May}, conditioning of the optimization problem \cite{Jerez2011}, model reduction \cite{Hovland2006}, (approximations of) explicit MPC \cite{Bemporad2002, Bemporad2011, Kvasnica2011Dec, Genuit2011}, and constraint removal techniques \cite{Jost2013,Jost2015,Dyrska2023Jun,Nie2019Jun,Nouwens2021b,Nouwens2023}.

In this paper, we are particularly interested in online constraint removal techniques. These techniques remove, at each time step, a subset of the constraints to simplify the optimization problem (and thus speed up computations). Typically, online constraint removal techniques can remove more constraints from the optimization problem compared to offline methods \cite{Paulraj2010,Roald2019}, as online methods can exploit specific ``local'' properties of the optimization problem, for instance, regions of activity \cite{Jost2013}, contraction properties of the cost function \cite{Jost2015, Dyrska2023Jun}, the refinement of coarse solutions \cite{Nie2019Jun}, reachable sets \cite{Nouwens2021b, Nouwens2023}, and the availability of a feasible input sequence \cite{Nouwens2023}.

In our previous works \cite{Nouwens2021b, Nouwens2023}, we proposed a constraint-adaptive MPC (ca-MPC) framework that can be used to detect and remove constraints from the optimization problem, without changing the minimizer and thus with identical closed-loop behavior. Our framework exploits, amongst other information, feasible input sequences. Obtaining such sequences is natural in an MPC setting through, for example, extending the optimal solution from the previous time step based on the available terminal ingredients. Extending the previous optimal solution is often used to guarantee stability and recursive feasibility of the MPC feedback law. However, some MPC setups lack the formal terminal ingredients and use soft-constraints to guarantee feasibility of the optimization problem. For soft-constrained MPC setups the existing constraint removal strategies are not highly effective. Hereto, in this work, we utilize the ideas from \cite{Nouwens2023} and propose a new computationally efficient constraint removal for soft-constrained MPC based on feasible input sequences. Crucially, for soft-constrained MPC, it will be trivial to generate feasible input sequences. We will demonstrate the method using an offset-free output tracking example, as this settings turns out to be particularly well-suited for our soft-constrained ca-MPC extension. 

The remainder of this paper is structured as follows. We will start with the preliminaries in Section~\ref{sec:preliminaries}. Second, in Section~\ref{sec:method}, we will present the soft-constrained ca-MPC scheme, after which we apply it to an offset-free output tracking MPC setup. Fourth, in Section~\ref{sec:numerical}, we demonstrate our method using a two-dimensional thermal regulation example containing 2030 inequality constraints. Finally, we state the conclusions in Section~\ref{sec:conclusion}. 


\vspace{-0.1em}
\section{Preliminaries}\label{sec:preliminaries}
In this section, we will introduce the system to be controlled, the MPC setup, and results for ellipsoidal sets.

\subsection{System description}
In this paper, we consider systems described by the linear time-invariant (LTI) discrete-time model
\begin{subequations}\label{eq:CDC_ss_model}
	\begin{align}
		\bm{x}_{k+1} &= \bm{Ax}_k+\bm{Bu}_k,\\
		\bm{y}_{k} &= \bm{Cx}_k + \bm{Du}_k.
	\end{align}
\end{subequations}
Here, $\bm{x}_k\in\mathbb{R}^{n_x}$, $\bm{u}_k\in\mathbb{R}^{n_u}$, and $\bm{y}_k\in\mathbb{R}^{n_y}$, denote the state, input, and output, respectively at time $k\in\mathbb{N}$. Moreover, $\bm{A},\ \bm{B},\ \bm{C},\ \bm{D}$ denote matrices of appropriate dimensions.

\subsection{MPC setup}
We will consider a MPC feedback law for an LTI model subject to linear inequality constraints and a quadratic cost function. By eliminating the dynamics, we obtain the following generic condensed MPC setup that is compatible with many typical MPC formulations with soft constraints \cite{Jerez2011, kerrigan2000soft}:
\begin{subequations}\label{eq:CDC_MPC_condensed}
	\begin{align}
		\underset{\bm{v},\ \bm{\varepsilon}}{\text{minimize}} \quad& \frac{1}{2}\bm{v}^\top\bm{H}\bm{v} + \bm{v}^\top\bm{F}\bm{z}_k + \bm{\rho}^\top\bm{\varepsilon}, \label{eq:CDC_MPC_condensed_a}\\
		\text{subject to}\quad &\bm{Wv} \leq \bm{c} + \bm{L}\bm{z}_k + \bm{\varepsilon}, \label{eq:CDC_MPC_condensed_b}\\
		&\bm{0}_{n_c}\leq\bm{\varepsilon}. \label{eq:CDC_MPC_condensed_c}
	\end{align}
\end{subequations}
Here, $\bm{v}\in\mathbb{R}^{n_v}$, $\bm{z}_k\in\mathbb{R}^{n_z}$, and $\bm{\varepsilon}\in\mathbb{R}^{n_c}$ denote the sequence of predicted control inputs, the initial state and possibly reference information at time $k$, and slack variables, respectively. The matrices $\bm{H}\in\mathbb{R}^{n_v\times n_v}_{\succ0}$ and $\bm{F}\in\mathbb{R}^{n_v\times n_z}$ define the quadratic part of the cost function and $\bm{\rho}\in\mathbb{R}^{n_c}_{>0}$ defines the penalty corresponding to a constraint violation. The matrices $\bm{W}\in\mathbb{R}^{n_c\times n_v}$, $\bm{L}^{n_c\times n_z}$ and vector $\bm{c}\in\mathbb{R}^{n_c}$ parameterize the soft-constrained inequality constraints and $\bm{0}_{n_c}$ is the zero vector of length $n_c$ (when we omit the subscript, we assume that the dimensions are clear from context). Note that there always exists a $\bm{\varepsilon}\geq\bm{0}$ sufficiently large such that a feasible solution to \eqref{eq:CDC_MPC_condensed} exists.

\begin{remark}
	In this paper, we consider an MPC setup where all constraints are soft constraints. Extensions to MPC setups where a subset of the constraints are soft constraints is straightforward. In such cases, it is needed to generate input sequences that are feasible with respect to the hard constraints. An easy scenario, where this is the case, is when all hard constraints are input constraints. 
\end{remark}

\subsection{Reduced MPC setup}
As we aim to remove, at each time step, a subset of the constraints from the MPC problem \eqref{eq:CDC_MPC_condensed}, we introduce the reduced MPC setup given by
\begin{subequations}\label{eq:CDC_MPC_condensed_red}
	\begin{align}
		\underset{\bm{v},\ \bm{\varepsilon}_\mathbb{A}}{\text{minimize}} \quad& \frac{1}{2}\bm{v}^\top\bm{H}\bm{v} + \bm{v}^\top\bm{F}\bm{z}_k + \bm{\rho}_\mathbb{A}^\top\bm{\varepsilon}_\mathbb{A},\\
		\text{subject to}\quad &\bm{W}_\mathbb{A}\bm{v} \leq \bm{c}_\mathbb{A} + \bm{L}_\mathbb{A}\bm{z}_k + \bm{\varepsilon}_\mathbb{A},\\
		&\bm{0}_{n_r}\leq\bm{\varepsilon}_\mathbb{A}.
	\end{align}
\end{subequations}
Here, we use the notation $\bm{W}_\mathbb{A}\in\mathbb{R}^{n_r\times n_v}$ to denote the matrix that consists of the rows in $\bm{W}$ that have a row index in $\mathbb{A}\subseteq\mathbb{N}_{[1,n_c]}=\{1,2,\cdots,n_c\}$. This is similar for $\bm{c}_\mathbb{A}$, $\bm{W}_\mathbb{A}$, $\bm{L}_\mathbb{A}$, $\bm{\varepsilon}_\mathbb{A}$. Note that $\bm{\varepsilon}_\mathbb{A}$ implies that the number of decision variables is decreased as well.

As mentioned previously, we will adapt the constraints at each time step and, in particular, we will base this on the initial state $\bm{z}_k$ and a feasible input sequence $\tilde{\bm{v}}_k$, which is assumed to be available at time $k$. More formally, given $\tilde{\bm{v}}_k\in\mathbb{R}^{n_v}$, we aim to compute $\mathbb{A}(\bm{z}_k,\tilde{\bm{v}}_k)\subseteq\mathbb{N}_{[1,n_c]}$ such that the minimizers of \eqref{eq:CDC_MPC_condensed} and \eqref{eq:CDC_MPC_condensed_red} are identical.

\subsection{Ellipsoidal set}
The ca-MPC method uses ellipsoidal bounds defined as
\begin{align}
	\mathcal{E}(\bm{P},\bm{q}) := \{\bm{x}\in\mathbb{R}^n\mid\|\bm{P}(\bm{x}-\bm{q})\|_2\leq 1\}\subset\mathbb{R}^{n_v}.
\end{align}
Here, $\bm{P}\in\mathbb{R}^{n_v\times n_v}$ is an invertible (possibly indefinite) matrix that controls the shape and size of the ellipsoid, and $\bm{q}\in\mathbb{R}^{n_v}$ is a vector that represents the center of the ellipsoid. 

\subsection{Ellipsoidal set and half-space intersection}\label{sec:intersection}
In the proposed algorithm below, we will determine which constraint index is included in $\mathbb{A}$ by evaluating if an ellipsoid $\mathcal{E}(\bm{P},\bm{q})$ is entirely contained in the half-space defined by an inequality constraint $\{\bm{w}\mid\bm{wv}\leq c\}$. To this end, we introduce the ellipsoid half-space intersection check,
\begin{align}\label{eq:CDC_intersection_def}
	\|\bm{w}\bm{P}^{-1}\|_2\leq |c - \bm{wq}|.
\end{align}
If \eqref{eq:CDC_intersection_def} holds, then $\mathcal{E}(\bm{P},\bm{q})\cap\{\bm{w}\mid\bm{wv}\leq c\} = \mathcal{E}(\bm{P},\bm{q})$, assuming $\mathcal{E}(\bm{P},\bm{q})\cap\{\bm{w}\mid\bm{wv}\leq c\} \neq \emptyset$. This assumption is satisfied when $\tilde{\bm{v}}_k$ does not violate any constraint \cite{Nouwens2023}.

\section{soft-constrained ca-MPC}\label{sec:method}
In this section, we will present the soft-constrained ca-MPC scheme. To this end, we will start by briefly summarizing one of the key insights from \cite{Nouwens2023}, where a feasible input sequence is used to bound the minimizer of a hard-constrained MPC problem. \emph{A minimizer of a convex optimization problem satisfies first-order optimality conditions, i.e, the negative gradient of the cost function is an element of the normal cone of the constraint set \cite{Nocedal2006}. Hence, outer approximations of the normal cone, based on a feasible input sequence, can be used to compute an ellipsoidal bound on the constrained minimizer}. We will, similar to this key idea in \cite{Nouwens2023}, use first-order optimality conditions to detect and remove constraints. In addition, we will exploit a key feature of soft constraints, i.e., it is trivial to generate feasible solutions for \eqref{eq:CDC_MPC_condensed}.

\begin{remark}
	The presented approach can be extended to include the reachability analysis from \cite{Nouwens2023} as well. We did not do this here, as we specifically focus on first-order optimality as this proved to be effective and requires minimal effort from the user to implement.
\end{remark}

The first step in our ca-MPC algorithm is to compute a feasible solution $(\tilde{\bm{v}}_k,\tilde{\bm{\varepsilon}}_k)$ to \eqref{eq:CDC_MPC_condensed}. To construct this solution, we start with $\tilde{\bm{v}}_k\in\mathbb{R}^{n_v}$ as our predicted input sequence at time $k$. Then, we ensure feasibility by computing $\tilde{\bm{\varepsilon}}_k$ as 
\begin{align}
	\tilde{\bm{\varepsilon}}_k:=\max(0,\bm{W}\tilde{\bm{v}}_k - \bm{c} - \bm{L}\bm{z}_k).
\end{align}
Trivially, $\tilde{\bm{\varepsilon}}_k = \bm{0}_{n_c}$ when $\tilde{\bm{v}}_k$ satisfies $\bm{W}\tilde{\bm{v}}_k \leq \bm{c} + \bm{L}\bm{z}_k$.

Based on the pair $(\tilde{\bm{v}}_k,\tilde{\bm{\varepsilon}}_k)$, we seek an ellipsoidal bound on the minimizer $\bm{v}^\star_k\in\mathcal{E}(\bm{P},\bm{q})$. This bound is then used to remove constraints from the optimization problem using the result from Section~\ref{sec:intersection}. Bounding only $\bm{v}^\star_k$ instead of both $\bm{v}^\star_k$ and $\bm{\varepsilon}^\star$ is the key insight that enables the soft-constrained ca-MPC scheme. 

To obtain the ellipsoid bounding $\bm{v}^\star_k$, we compute the first-order optimality conditions for the soft-constrained MPC problem \eqref{eq:CDC_MPC_condensed}, which yields
\begin{subequations}\label{eq:CDC_first_order_optimality}
\begin{align}
	-\begin{bmatrix}
		\bm{H}\bm{v}^{\star}_k+\bm{F}\bm{z}_k \\ \bm{\rho}
	\end{bmatrix} \in \mathcal{N}(\bm{v}^\star_k,\bm{\varepsilon}^\star_k),
\end{align}
where the normal cone $\mathcal{N}(\bm{v}^\star_k,\bm{\varepsilon}^\star_k)\subseteq\mathbb{R}^{n_v+n_c}$ is given by
\begin{align}\nonumber
	\mathcal{N}(\bm{v}^\star_k,\bm{\varepsilon}^\star_k) = \{&\bm{p}\in\mathbb{R}^{n_v+n_c}\mid \bm{p}^\top\left(\begin{bmatrix}\bm{v}\\\bm{\varepsilon}\end{bmatrix} - \begin{bmatrix}\bm{v}^\star_k\\\bm{\varepsilon}^\star_k\end{bmatrix}\right)\leq 0,\\ \label{eq:CDC_normalcone_b}
	&\text{for all } \bm{v},\ \bm{\varepsilon}\ \text{subject to}\ \eqref{eq:CDC_MPC_condensed_b},\ \eqref{eq:CDC_MPC_condensed_c}\}.
\end{align}
\end{subequations}
Instead of evaluating \eqref{eq:CDC_normalcone_b} at all pairs $(\bm{v},\bm{\varepsilon})$ subject to \eqref{eq:CDC_MPC_condensed_b}, \eqref{eq:CDC_MPC_condensed_c}, we outer approximate the normal cone by evaluating \eqref{eq:CDC_normalcone_b} at $(\tilde{\bm{v}}_k,\tilde{\bm{\varepsilon}}_k)$. By evaluating \eqref{eq:CDC_normalcone_b} for a single pair, we obtain the following necessary condition for the minimizer:
\begin{align}\label{eq:CDC_necessary_condition}
	-\begin{bmatrix}
		\bm{H}\bm{v}^{\star}_k+ \bm{F}\bm{z}_k \\ \bm{\rho}
	\end{bmatrix}^\top\left(\begin{bmatrix}\tilde{\bm{v}}_k\\\tilde{\bm{\varepsilon}}_k\end{bmatrix} - \begin{bmatrix}\bm{v}^\star\\\bm{\varepsilon}^\star_k\end{bmatrix}\right)\leq 0.
\end{align}
After expanding and collecting the terms, we obtain
\begin{align}\label{eq:CDC_ellpise_v_1}
	(\bm{Hv}^\star_k+\bm{Fz}_k)^\top(\bm{v}^\star_k-\tilde{\bm{v}}_k) \leq \bm{\rho}^\top(\tilde{\bm{\varepsilon}}_k - \bm{\varepsilon}^\star_k).
\end{align}
The inequality \eqref{eq:CDC_ellpise_v_1} denotes an ellipse in $\bm{v}^\star_k$ that scales with $\bm{\rho}^\top\bm{\varepsilon}^\star_k$. To obtain an ellipsoidal bound on $\bm{v}_k^\star$, that does not depend on $\bm{\varepsilon}_k^\star$, we use the observation that $0\leq\bm{\rho}^\top\bm{\varepsilon}_k^\star$, see \eqref{eq:CDC_MPC_condensed_c}. This observation enables the upper bound $\bm{\rho}^\top(\tilde{\bm{\varepsilon}}_k-\bm{\varepsilon}^\star_k) \leq \bm{\rho}^\top\tilde{\bm{\varepsilon}}_k$, which leads to the bound on $\bm{v}^\star_k$,
\begin{align}\label{eq:CDC_ellpise_v_2}
	(\bm{Hv}^\star_k+\bm{Fz}_k)^\top(\bm{v}^\star_k-\tilde{\bm{v}}_k) \leq \bm{\rho}^\top\tilde{\bm{\varepsilon}}_k.
\end{align}
Which is equivalent to $\bm{v}^\star_k\in\mathcal{E}(\bm{P},\bm{q})$, with
\begin{subequations}\label{eq:CDC_def_Pq}
	\begin{align} 
		\bm{P} &= \frac{1}{\sqrt{\sigma}} \bm{G},\ 
		\bm{q} = \frac{1}{2}(\tilde{\bm{v}}_k - \bm{H}^{-1}\bm{F}\bm{z}_k),\ \bm{H} = \bm{G}^\top\bm{G},\\
		\sigma &= \bm{\rho}^\top\tilde{\bm{\varepsilon}}_k + \frac{1}{4}\|\bm{G}(\tilde{\bm{v}}_k + \bm{H}^{-1}\bm{Fz}_k)\|_2^2.
	\end{align}
\end{subequations}
Based on \eqref{eq:CDC_def_Pq}, we observe that $\bm{v}^\star_k$ is bounded by an ellipse centered at the mean between $-\bm{H}^{-1}\bm{F}\bm{z}_k$ and $\tilde{\bm{v}}_k$, and inflated based on the degree of infeasibility and distance between $\tilde{\bm{v}}_k$ and $-\bm{H}^{-1}\bm{F}\bm{z}_k$.
\begin{remark}\label{rem:CDC_rem1}
	The expression for $\sigma$ provides a guideline on how we should choose $\tilde{\bm{v}}_k$, as, typically, we want $\sigma$ to be small. Hence, on one hand we want to be close to the unconstrained minimizer to reduce $\|\bm{G}(\tilde{\bm{v}}_k + \bm{H}^{-1}\bm{Fz}_k)\|_2^2$, while on the other hand we would like to reduce the infeasibility of our prediction to make $\bm{\rho}^\top\tilde{\bm{\varepsilon}}_k$ small. Therefore, there is a balancing act between reducing infeasibility and remaining close to the unconstrained minimizer. Of course, when the unconstrained minimizer is feasible, we have trivially solved the optimization problem.
\end{remark}

Based on the ellipsoidal bound, we use the inequality from Section~\ref{sec:intersection} to define the set-valued mapping $\mathbb{A}(\bm{z}_k,\tilde{\bm{v}}_k)$,
\begin{align}\label{eq:CDC_Aset_def}
	\mathbb{A}(\bm{z}_k,\tilde{\bm{v}}_k) &= \{j\in\mathbb{N}_{[1,n_c]} \mid\sqrt{\sigma}\|\bm{W}_j\bm{G}^{-1}\|_2 >\\ \nonumber
	&\hspace{8em}|\bm{c}_j + \bm{L}_j\bm{z}_k - \bm{W}_j\bm{q}|\}.
\end{align}
Here, we use $\bm{W}_j$ to denote the $j$-th row from $\bm{W}$ (the same holds for $\bm{c}_j$ and $\bm{L}_j$). An interesting observation is that \eqref{eq:CDC_Aset_def} does not explicitly specify that the constraints violated by $\tilde{\bm{v}}_k$ are included in $\mathbb{A}$. As it turns out, depending on the design of $\bm{\rho}$, e.g., $\bm{\rho}$ sufficiently large \cite{kerrigan2000soft}, these constraints are automatically included by \eqref{eq:CDC_Aset_def}. However, when $\bm{\rho}$ is not sufficiently large, we may encounter the situation where all elements of $\mathcal{E}(\bm{P},\bm{q})$ violate a particular constraint. This violates the assumption that the intersection between the ellipsoid and half-space for a particular inequality constraint is non-empty, see Section~\ref{sec:intersection}. To alleviate this problem and avoid complex design rules on $\bm{\rho}$, we modify \eqref{eq:CDC_Aset_def} by including all constraints that are violated by $\tilde{\bm{v}}_k$,
\begin{align}\label{eq:CDC_Aset_def2}
	\mathbb{A}(\bm{z}_k,\tilde{\bm{v}}_k) = &\{j\in\mathbb{N}_{[1,n_c]} \mid\sqrt{\sigma}\|\bm{W}_j\bm{G}^{-1}\|_2 >\\ \nonumber
	&\hspace{3em}|\bm{c}_j + \bm{L}_j\bm{z}_k - \bm{W}_j\bm{q}|\ \text{or}\ \tilde{\bm{\varepsilon}}_{j,k}>0\},
\end{align}
where $\tilde{\bm{\varepsilon}}_{j,k}$ denotes the $j$-th element of $\tilde{\bm{\varepsilon}}_k$. 

Observe that a significant fraction of \eqref{eq:CDC_Aset_def2} can be either pre-computed, e.g., $\|\bm{W}_j\bm{G}^{-1}\|_2$, or is already required to set up the MPC problem, e.g., $\bm{F}\bm{z}_k$ and $\bm{c}_j + \bm{L}_j\bm{z}_k$, so, these do not lead to additional computational costs. The number of operations specific to the ca-MPC framework is $2n_c(n_v+1) + 3n_v^2 + n_v + 1$, which, crucially, scales linear in $n_c$. 

Last, we provide an overview of the method in Algorithm~\ref{alg:ca-MPC_algorithm}. Note that the only ca-MPC specific offline setup is pre-computing $\|\bm{W}_j\bm{G}^{-1}\|_2$ in lines 2-5. In line 7, we measure $\bm{z}_k$, which is, for example, a stacked vector containing the current state, previous input, and information regarding a reference trajectory. In lines 8-11, we choose $\tilde{\bm{v}}_k$ based on the shifted previous optimal minimizer $\bm{v}_{[n_u+1:Nn_u],k-1}^{\star}$, where the subscript $[n_u+1:Nn_u],k-1$ denotes that we select the $(n_u+1)$-th up to $Nn_u$-th elements at time $k-1$. When the minimizer at time $k-1$ is not available, e.g., at time $k=0$, we use the unconstrained minimizer $-\bm{H}^{-1}\bm{Fz}_k$. Recall that the method allows all $\tilde{\bm{v}}_k\in\mathbb{R}^{n_v}$, but certain choices show better constraint removal properties, see also Remark~\ref{rem:CDC_rem1} above. In line 20, we extract the rows from the constraint matrices and vectors based on $\mathbb{A}$; this is required to build the reduced MPC problem. In lines 21-23, we solve the quadratic program, compute the optimal input based on the minimizer, and apply the input to the plant. Note that $\bm{v}_k$ does not necessarily contain the input, e.g., when we optimize over delta's of the input relative to $\bm{u}_{k-1}$.


\section{Offset-free output tracking MPC}\label{sec:offset_free}
The soft-constrained ca-MPC scheme described in the previous section applies to all MPC setups that can be captured in the formulation \eqref{eq:CDC_MPC_condensed}, which includes a large class of control problems for linear plant models. In this section, we will focus on an offset-free output tracking problem as it is both an industry-relevant MPC setup and particularly well-suited to our soft-constrained ca-MPC scheme. 

The offset-free output tracking MPC setup is given by the optimization problem \cite{Pannocchia}
\begin{subequations}\label{eq:CDC_general_MPC}
	\begin{align}
			\hspace{-0.5em}\underset{}{\text{minimize}}\ \quad& \sum_{i=1}^{N} \|\bm{Cx}_{i|k}-\bm{y}^\text{ref}_{k+i}\|^2_{\bm{Q}} + \|\bm{\Delta u}_{i-1|k}\|^2_{\bm{R}}		,\label{eq:CDC_general_MPC_a}\\\nonumber &\hspace{1em} + \bm{\rho}_x^\top\bm{\varepsilon}_{x,i} + \bm{\rho}_u^\top\bm{\varepsilon}_{u,i-1} + \bm{\rho}_\Delta^\top\bm{\varepsilon}_{\Delta,i-1},\\
			\hspace{-0.5em}\text{subject to}\quad&\bm{x}_{i+1|k} = \bm{Ax}_{i|k} + \bm{Bu}_{i|k},\ i\in\mathbb{N}_{[0,N-1]},\\
			&\bm{u}_{i|k} = \bm{u}_{i-1|k} + \bm{\Delta u}_{i|k},\ i\in\mathbb{N}_{[0,N-1]},\\
			&\bm{x}_{0|k} = \bm{x}_k,\ \bm{u}_{-1|k} = \bm{u}_{k-1},\\
			&\bm{M}_x\bm{x}_{i|k}\leq \bm{g}_x + \bm{\varepsilon}_{x,i},\ i\in\mathbb{N}_{[1,N]},\\ 
			&\bm{M}_u\bm{u}_{i|k}\leq \bm{g}_u + \bm{\varepsilon}_{u,i},\ i\in\mathbb{N}_{[0,N-1]},\\
			&\bm{M}_\Delta\bm{\Delta u}_{i|k} \leq \bm{g}_\Delta + \bm{\varepsilon}_{\Delta,i},\ i\in\mathbb{N}_{[0,N-1]},\\
			&\bm{0}\leq\bm{\varepsilon}_{x,i},\  i\in\mathbb{N}_{[1,N]},\\
			&\bm{0}\leq\bm{\varepsilon}_{u,i},\ \bm{0}\leq\bm{\varepsilon}_{\Delta,i},\  i\in\mathbb{N}_{[0,N-1]}.
		\end{align}
\end{subequations}
Here, we use $\|\bm{\Delta u}\|^2_{\bm{R}}$ to denote $\bm{\Delta u}^\top\bm{R}\bm{\Delta u}$ and the $i|k$ subscripts denotes the $i$-th prediction step made at time $k$. In \eqref{eq:CDC_general_MPC}, matrices $\bm{Q}\in\mathbb{R}^{n_y\times n_y}$ and $\bm{R}\in\mathbb{R}^{n_u\times n_u}$ denote a positive semi-definite and a positive definite matrix, respectively. The vector $\bm{y}_k^\text{ref}\in\mathbb{R}^{n_y}$ denotes the reference and the matrix $\bm{M}$ and vector $\bm{g}$ define the inequality constraints for, respectively, the state, input, and delta input, depending on the subscript. We optimize \eqref{eq:CDC_general_MPC} with respect to $\bm{x}_{i|k}$, $\bm{u}_{i-1|k}$, $\bm{\Delta u}_{i-1|k}$, $\bm{\varepsilon}_{x,i}$, $\bm{\varepsilon}_{u,i-1}$, and $\bm{\varepsilon}_{\Delta,i-1}$ for $i\in\mathbb{N}_{[1,N]}$.

To transform \eqref{eq:CDC_general_MPC} into \eqref{eq:CDC_MPC_condensed}, we define 
\begin{subequations}
	\begin{align}
	\bm{z}_k = &[\bm{x}_k^\top\ \bm{u}_{k-1}^\top\ \bm{y}_{k+1}^{\text{ref}\top}\ \cdots\ \bm{y}_{k+N}^{\text{ref}\top}]^\top,\\
	\bm{v}_k = &[\bm{\Delta u}_{0|k}^\top\ \cdots\ \bm{\Delta u}_{N-1|k}^\top]^\top,\\
	\bm{\varepsilon} = &[\bm{\varepsilon}_{x,1}^\top\ \bm{\varepsilon}_{u,0}^\top\ \bm{\varepsilon}_{\Delta,0}^\top\ \cdots\ \bm{\varepsilon}_{x,N}^\top\ \bm{\varepsilon}_{u,N-1}^\top\ \bm{\varepsilon}_{\Delta,N-1}^\top]^\top,
	\end{align}
\end{subequations}
after which transformations that condense the MPC problem can be applied to obtain $\bm{H}$, $\bm{G}$, $\bm{F}$, $\bm{W}$, $\bm{c}$, $\bm{L}$, and $\bm{\rho}$.

Based on the MPC formulation \eqref{eq:CDC_general_MPC}, we can analyze why \eqref{eq:CDC_general_MPC} is well-suited to the soft-constrained ca-MPC scheme. First of all, the delta-input enables a particularly good and easy-to-compute input sequence, namely, $\tilde{\bm{v}}_k = [\bm{\Delta u}_{1|k-1}^{\star\top}\ \cdots\ \bm{\Delta u}_{N-1|k-1}^{\star\top}\ \bm{0}_{n_u}^\top]^\top$. Indeed, we can simply extend the previous optimal solution with a zero input, which is the steady-state solution for a constant reference. Moreover, recall that the size of the ellipsoid bounding the minimizers is determined, in part, by $\bm{\rho}^\top\tilde{\bm{\varepsilon}}_k$. The particular choice of $\tilde{\bm{v}}_k$ is expected to result in small constraint violations, i.e., we expect $\bm{\rho}^\top\tilde{\bm{\varepsilon}}_k$ to be small. 

The second reason why \eqref{eq:CDC_general_MPC} is particularly well suited to the soft-constrained ca-MPC scheme, is that the objective is to track the reference well. In other words, we can expect to be ``close" to the reference in some sense. To see why this is valuable, recall that the ellipsoid $\mathcal{E}(\bm{P},\bm{q})$ also scales based on the distance between the unconstrained minimizer $-\bm{H}^{-1}\bm{Fz}_k$, and the predicted input sequence $\tilde{\bm{v}}_k$. This means that depending on the output reference, the unconstrained minimizer is expected to be close to the predicted input sequence. For example, when moving between two set points, the unconstrained minimizer will be closer to $\tilde{\bm{v}}_k$, when we ramp between them as opposed to instantaneously switching. Hence, we expect to remove more constraints when we have an appropriate reference that connects the set points. We will demonstrate the effectiveness of the constraint removal method on an offset-free output tracking problem in the next section.

\floatstyle{spaceruled}
\restylefloat{algorithm}
\begin{algorithm}\small
	\caption{The soft-constrained ca-MPC scheme}
	\begin{algorithmic}[1]
		\State $k\leftarrow0,\ \bm{H}\in\mathbb{R}^{n_v\times n_v}_{\succ0},\ \bm{G}\in\mathbb{R}^{n_v\times n_v},\ \bm{F}\in\mathbb{R}^{n_v\times n_z},\newline \bm{W}\in\mathbb{R}^{n_c\times n_v},\ \bm{L}\in\mathbb{R}^{n_c\times n_z},\ \bm{c}\in\mathbb{R}^{n_c},\ \bm{\rho}\in\mathbb{R}^{n_c}_{>0},\ j\leftarrow 1$
		\For{$j\leq n_c$}
		\State $\zeta_j \leftarrow \|\bm{W}_j\bm{G}^{-1}\|_2$
		\State $j \leftarrow j + 1$
		\EndFor
		\While{\texttt{true}}
		\State \textsc{Measure } $\bm{z}_k$
		\If{$k=0$}
		\State $\tilde{\bm{v}}_k \leftarrow -\bm{H}^{-1}\bm{Fz}_k$
		\Else
		\State $\tilde{\bm{v}}_k \leftarrow [\bm{v}_{[n_u+1:Nn_u],k-1}^{\star\top}\ \bm{0}_{n_u}^\top]^\top$
		\EndIf
		\State $\tilde{\bm{\varepsilon}}_k \leftarrow \max(0,\bm{W}\tilde{\bm{v}}_k - \bm{c} - \bm{Lz}_k)$
		\State $\bm{q}\leftarrow\frac{1}{2}(\tilde{\bm{v}}_k - \bm{H}^{-1}\bm{Fz}_k)$
		\State $\sigma \leftarrow \bm{\rho}^\top\tilde{\bm{\varepsilon}}_k + \frac{1}{4}\|\bm{G}(\tilde{\bm{v}}_k + \bm{H}^{-1}\bm{Fz}_k)\|_2^2$
		\State $j\leftarrow1,\ \mathbb{A}\leftarrow\emptyset$
		\For{$j\leq n_c$}
		\If{$\sqrt{\sigma}\zeta_j > |\bm{c}_j + \bm{L}_j\bm{z}_k - \bm{W}_j\bm{q}|$ \texttt{or} $\tilde{\bm{\varepsilon}}_{j,k}>0$}
		\State $\mathbb{A} \leftarrow \mathbb{A} \cup \{j\}$
		\EndIf	
		\State $j \leftarrow j+1$					
		\EndFor
		\State \textsc{Extract } $\bm{W}_\mathbb{A},\ \bm{L}_\mathbb{A},\ \bm{c}_\mathbb{A},\ \bm{\rho}_\mathbb{A}$
		\State $(\bm{v}_k^\star,\bm{\varepsilon}^\star_\mathbb{A}) \leftarrow\ $\textsc{solve \eqref{eq:CDC_MPC_condensed_red}}
		\State \textsc{Extract } $\bm{u}_k \leftarrow (\bm{v}_k^\star, \bm{z}_k)$
		\State \textsc{Apply $\bm{u}_k$ to plant}
		\State $k=k+1$
		\EndWhile
	\end{algorithmic}\label{alg:ca-MPC_algorithm}
\end{algorithm}


\section{Numerical case study}\label{sec:numerical}
In this section, we will demonstrate the soft-constrained ca-MPC scheme using a two-dimensional thermal regulation example. Here, we aim to heat a particular region without exceeding temperature upper bounds on our entire two-dimensional domain, see Figure~\ref{fig:CandM}. This type of problem is relevant for hyperthermia-enhanced cancer treatments, see, e.g., \cite{Deenen2020Dec}. We will start by defining our system and MPC setup. Hereafter, we will compare the computational complexity of our ca-MPC scheme and the original MPC. 

\subsection{System definition and MPC setup}
\begin{figure}[th!]
	\centering
	\includegraphics[width=8.4cm]{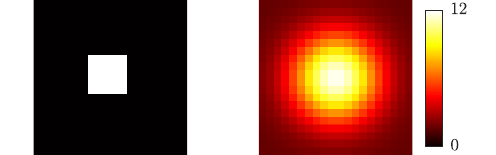}
	\caption{Left: the union between the black and white pixels denote the domain $\Omega$ and the white pixels denote the outputs. Right: the (Gaussian) temperature upper bound.}
	\label{fig:CandM}
\end{figure}

We consider a thermal system modeled by the following partial differential equation
\begin{subequations}\label{eq:CDC_thermal_pde}
	\begin{align}
		&\dot{T}(r,t) = \alpha\nabla^2T(r,t) + \beta T(r,t) + \textstyle\sum_{j=1}^{3} \gamma_j u_j(t),
	\end{align}
	for $r\in\Omega$, with boundary condition
	\begin{align}
		&\alpha\nabla T(r,t)\cdot \bm{n} = T(r,t),\ r\in\partial\Omega.
	\end{align}
\end{subequations}
Here, $\Omega=[0,1]\times[0,1]$, $T:\Omega\times\mathbb{R}_{\geq0} \rightarrow \mathbb{R}$, $\alpha=2.5\cdot10^{-4}$, $\beta=2\cdot10^{-2}$, $\gamma_j:\Omega\rightarrow\mathbb{R}_{\geq0}$, $u_j:\mathbb{R}_{\geq0}\rightarrow\mathbb{R}_{\geq0}$, and $\bm{n}\in\mathbb{R}^2$, denote the domain, temperature, diffusivity, damping, distributed heat load, control input, and outward facing surface normal, respectively. As mentioned, we constrain the temperature below an upper bound, as given by  $T(r,t)\leq\bar{T}(r)$ for $r\in\Omega$, see Figure~\ref{fig:CandM}. 

\begin{figure}[th!]
	\centering
	\includegraphics[width=8.4cm]{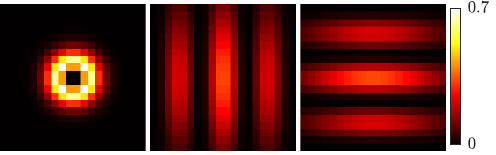}
	\caption{The spatially distributed heat loads $\gamma_1,\ \gamma_2$, and $\gamma_3$ (from left to right).}
	\label{fig:B}
\end{figure}

After spatially and temporally discretizing \eqref{eq:CDC_thermal_pde} on a square $20\times20$ grid with a sample time of 1 second, we obtain a discrete-time state-space model \eqref{eq:CDC_ss_model} with $n_x=400$ states and $n_y = 25$ outputs, again, see Figure~\ref{fig:CandM}. To obtain an offset-free MPC scheme, we write the input in incremental form: $\bm{u}_{k+1} = \bm{u}_{k} + \Delta\bm{u}_{k}$. The heat loads $\gamma_i$ on the discrete spatial grid are shown in Figure~\ref{fig:B}. Similarly, discretization of the temperature upper bound gives $\bm{M}_x = \bm{I}_{n_x}$ and $\bm{g}_x = \bm{\bar{T}}\in\mathbb{R}^{n_x}$, where $\bm{I}_{n_x}$ denotes the identity matrix of size $n_x$ and $\bar{\bm{T}}$ is shown in Figure~\ref{fig:CandM}. Additionally, we constrain our input to be non-negative and limited in power by $\bm{M}_u = \bm{I}_{n_u}\otimes \left[\begin{smallmatrix}1 \\ -1\end{smallmatrix}\right]$ and $\bm{g}_u = \bm{1}_m\otimes\left[\begin{smallmatrix}1 \\ 0\end{smallmatrix}\right]$, where $\otimes$ denotes the Kronecker product. The non-negative input in combination with the non-negative $\gamma_i$ ensure we cannot actively apply cooling, as is typical in hyperthermia cancer treatments. We choose $\bm{Q}=\bm{I}_{n_y}$, $\bm{R}=\bm{I}_{n_u}$, and $\bm{\rho} = \bm{1}_{n_c}$. Using $\bm{M}_x,\ \bm{g}_x,\ \bm{M}_u,\ \bm{g}_u,\ \bm{Q},\ \bm{R}$ and choosing a horizon $N=5$, we can transform the output tracking MPC \eqref{eq:CDC_general_MPC} to \eqref{eq:CDC_MPC_condensed} using the definitions from Section~\ref{sec:offset_free}. Note that our MPC scheme has a total of $n_c = 2030$ constraints and $n_v = 15$ control inputs. Recall that we obtain many constraints as the upper bound on the continuous spatial domain is discretized. We define the output reference as a linear ramp to the target temperature, $\bm{y}_k^\text{ref} = \bm{1}_{n_y}\min(10,10\frac{k}{30})$. We solve the resulting quadratic programs using a primal-dual interior point solver from the Matlab MPC toolbox \cite{MPC_Toolbox}.

\subsection{Results}
\begin{figure}[th!]
	\centering
	\includegraphics[width=8.4cm]{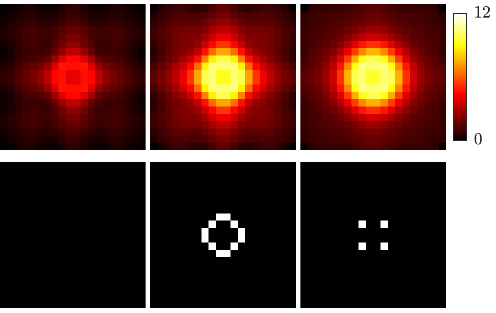}
	\caption{Top: the temperature at time steps 15, 30, and 45, respectively. Bottom: white pixels indicate that the corresponding constraint is added to $\mathbb{A}$ for at least one step in the horizon. Note that at time step 30, we approximately obtain the maximum number of constraints in the MPC problem.}
	\label{fig:XandC}
\end{figure}

\begin{figure}[th!]
	\vspace{1em}
	\centering
	\includegraphics[width=8.4cm]{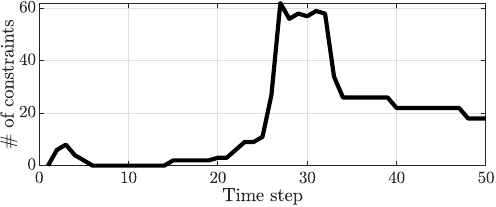}
	\caption{The number of constraints in the reduced MPC problem over time. The maximum number of constraints (62) is obtained at time step 27. Recall that the total number of constraints is 2030.}
	\label{fig:CoverT}
\end{figure}

In this section, we present the results comparing the soft-constrained ca-MPC scheme to the MPC setup \eqref{eq:CDC_MPC_condensed}. Recall that the ca-MPC setup has the exact same closed-loop behavior original soft-constrained MPC setup. Interestingly, when $\bm{\rho}$ is sufficiently large, the soft-constrained MPC scheme is equivalent to the corresponding hard-constrained MPC setup \cite{kerrigan2000soft}. First, in Figure~\ref{fig:XandC}, we show the temperature evolution over time for the ca-MPC scheme in combination with a visualization of the constraints that are included in the reduced MPC problem. As expected, the temperature in the center rises to follow the reference. However, crucially, only a small number of constraints at the discrete locations are considered in the reduced MPC problem. In Figure~\ref{fig:CoverT}, we show how the number of constraints evolves over time. Note that the number of constraints does not exceed 62, which is $3.1\%$ of the original number of constraints.

\begin{figure}[th!]
	\sbox0{\Cline{black}{solid}}\sbox1{\Cline{orange}{solid}}\sbox2{\Cline{blue}{solid}}\sbox3{\Cline{blue}{dashed}}
	\centering
	\includegraphics[width=8.4cm]{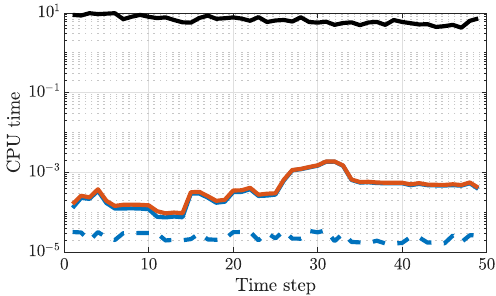}
	\caption{Computation time comparison between the original MPC setup \eqref{eq:CDC_MPC_condensed} (\usebox0), and the ca-MPC setup \eqref{eq:CDC_MPC_condensed_red} (\usebox1). The time spent by soft-constrained ca-MPC is broken down in all time requited to compute $\mathbb{A}$ (\usebox3) and solving the resulting quadratic program (\usebox2).}
	\label{fig:TvsT}
\end{figure}

The significant reduction in constraints is also reflected in the computation time, see Figure~\ref{fig:TvsT}. Here, our ca-MPC setup obtains a three orders of magnitude improvement in computation time compared to the original MPC setup. From Figure~\ref{fig:TvsT}, we observe that the time spent to compute $\mathbb{A}$ is negligible with respect to solving the resulting simplified quadratic program (QP), let alone compared to solving the original QP. This observation exemplifies the effectiveness and computational efficiency of the ca-MPC method. Indeed, for many MPC problems, determining which constraints can be excluded from the QP (computing $\mathbb{A}$) can be performed in a negligible amount of time. Recall that by using soft constraints, removing a constraint from the optimization problem simultaneously removes a decision variable from the QP, further simplifying the resulting ca-MPC scheme.



\section{Conclusion and outlook}\label{sec:conclusion}
In this paper, we presented an extension to ca-MPC that allows for online constraint removal using easy-to-compute input sequences for soft-constrained MPC. Crucially, our constraint removal scheme does not change the minimizer of the original MPC problem and, thus, the ca-MPC scheme inherits all stability and performance properties from the original MPC setup. The presented method uses an a priori generated input sequence to compute ellipsoidal bounds on the constrained minimizer to detect and remove inequality constraints that cannot be reached within the ellipsoidal bound. Similar to existing ca-MPC schemes, our extension has a low computational overhead, making it applicable to a large class of linear systems and MPC setups. We demonstrated the ca-MPC scheme on a two-dimensional thermal regulation problem with a temperature upper bound. The regular MPC setup had 2030 constraints, while our ca-MPC scheme only required a maximum of 62, while guaranteeing the same closed-loop performance. This resulted in a computational speed-up of three orders of magnitude.  

Exploiting the optimality properties of MPC proves to be effective at accelerating the computation of the online optimization problem. Future research interests include extensions to different cost functions and direct integration into specific optimization solvers, such as interior-point and active-set solvers. Other subjects of interest include different handling of the soft constraints, e.g., using one slack variable to bound all constraints and extensions to different slack variable penalties, such as mixed quadratic and linear penalties. Extensions to proximal point methods are also of interest. These methods solve a sequence of well-conditioned optimization problems, where each minimizer is close to the previous one. This property has the potential for highly effective constraint removal schemes.

\bibliographystyle{IEEEtran}
\bibliography{literature}

\end{document}